\documentstyle[pre,aps,psfig]{revtex}
\begin{document}

\date{\today}
\title{Interface pinning and slow ordering kinetics on infinitely ramified
fractal structures.}

\author{Umberto Marini Bettolo Marconi}
\address{Dipartimento di Matematica e Fisica, Universit\`a di Camerino,
         Via Madonna delle Carceri,I-62032 , Camerino, Italy}
\address{Istituto Nazionale di Fisica della Materia, Sez. di Camerino}         
\address{Istituto Nazionale di Fisica Nucleare, Sez. di Perugia}

\maketitle
\begin{abstract}
We investigate the time dependent Ginzburg-Landau 
(TDGL) equation for a
non conserved order parameter on
an infinitely ramified (deterministic) fractal lattice employing two
alternative methods:
the auxiliary field approach and 
a numerical method of integration of the equations of evolution.
In the first case the domain size evolves with time as $L(t)\sim t^{1/d_w}$,
where $d_w$ is the anomalous random walk exponent associated with 
the fractal and differs from the normal value $2$, which
characterizes all Euclidean lattices. Such a power law growth is identical
to the one observed in the study of the spherical model on the
same lattice, but fails to describe the asymptotic behavior
of the numerical solutions of the TDGL equation for 
a scalar order parameter.
In fact, the simulations performed on a two dimensional Sierpinski Carpet 
indicate that, after an initial stage dominated by a curvature reduction
mechanism \`a la Allen-Cahn, the system enters in a regime where the domain 
walls between competing phases are pinned by lattice defects.
 The lack of translational invariance determines a rough free energy 
landscape, the existence of many metastable minima and the suppression of
the marginally stable modes, which in translationally invariant systems
lead to power law growth and self similar patterns. On fractal structures
as the temperature vanishes the evolution is frozen, since
only thermally activated processes can sustain the growth of pinned domains.

\end{abstract}


\section*{introduction}

The study of the relaxation dynamics of a system initially in thermal 
equilibrium and abruptly rendered unstable by a sudden change of a 
controlling field has recently drawn considerable 
attention, not only because many physical properties may depend 
on the way a material reaches the equilibrium state, but also 
because it poses intriguing problems to the theory such as broken 
ergodicity, aging and dynamical scaling \cite{Bray},

After the quench,
i.e. a rapid lowering of the temperature below 
the critical point of a phase separating
system, the initial, disordered state looses
stability and the system undergoes a coarsening process, during which
the domains corresponding to different equilibrium phases 
compete to grow in magnitude. About twenty years ago,
Allen and Cahn \cite{Allen}  realized that when the order 
parameter is  non conserved,  
the driving force towards equilibrium stems from 
the tendency of the system to reduce the curvature of 
the domain walls and 
showed that the typical size  of the domains , $L$, 
increases in time with a power law
behavior $L(t) \sim t^{1/2}$.
In spite of the fact that the theory of phase 
ordering in homogeneous systems is 
pretty well understood, 
only recently  the ordering kinetics on
non-translationally invariant, fractal lattices,
has become a  subject of investigation.

The present author in collaboration with Petri \cite{UMBM}-
\cite{andalo} considered 
the role of deterministic fractal supports, with finite and infinite
order of ramification,
on the ordering process 
employing the so-called spherical model,
which has the advantage of rendering analytical approaches possible. 
It was found
by means of an explicit solution
that the spherical model on fractal lattices
of finite order of ramification, such as
the Sierpinski gasket of arbitrary embedding dimension
does not display a finite temperature phase transition;
on the contrary on Sierpinski Carpets, whose 
order of ramification is infinite
\cite{rami}-\cite{Gefen}, 
there exist an order-disorder transition provided that the spectral dimension 
$d_s$ exceeds the critical value $2$ \cite{dos}, in
accord with the Mermin-Wagner theorem \cite{Mermin}.

Interestingly, the study  
of the spherical model 
with non conserved order parameter
has revealed the existence, even on fractals, of
a characteristic length scale $L(t)$, which
increases in time in a power like fashion $L(t) \sim t^{1/z}$, and
of dynamical scaling for the correlation functions.
Such a dynamical exponent $z$ takes the value
$d_w=\ln(d+3)/\ln(2)$ on
Sierpinski gaskets of arbitrary embedding dimension 
$d$, and $d_w\simeq 2.10$ on the planar Sierpinski Carpet, where $d_w$ is
the random walk exponent.
These values of $z$ differ from
the Allen-Cahn universal value $z=2$, 
which characterizes the diffusive domain growth 
on standard lattices.
We have shown \cite{UMBM} that in order to fully 
characterize the static and dynamical
properties of the spherical model on fractal lattices two more
quantities are required: i) the fractal dimension
$d_f$ and ii) the spectral dimension $d_s$, which for many lattices
are related to $d_w$ via the Alexander-Orbach \cite{AO} relation:
$$
d_w=\frac{2 d_f}{d_s}.
$$

Unfortunately the analytically soluble 
spherical model does not yield predictions
which can be extrapolated to the physically interesting case 
of the scalar order parameter. 
In fact, the lack of sharp, well defined interfaces
between different phases, renders the spherical model
physically inadequate to describe 
the phase separation process of an Ising system.

 The purpose of the present paper is to investigate
the phase separation process on fractals with infinite ramification order, 
such as the Sierpinski Carpet family , for a non conserved
 scalar order parameter. We expect 
the latter to be very sensitive to the presence
of inhomogeneities  in contrast with  vector fields. 
The choice of a Sierpinski Carpet has been suggested by the fact that 
it represents perhaps the simplest example of non stochastic fractal
lattice, 
with infinite ramification order \cite{rami}, \cite{Frattali}.
In  section II we introduce the Ginzburg-Landau (GL)
model on the fractal lattice,
in III we consider an auxiliary field approach 
\cite{Desiena}
and discuss
its asymptotic behavior.
In IV we present numerical results
of the exact equations of motion and monitor in several ways the
growth process.
In V after stressing the similarities between the
auxiliary field method and the spherical model, we draw the
conclusions.

\section{Ginzburg-Landau model on a fractal lattice.}

We shall consider a scalar field $\phi_x$ whose properties depend on a
standard Ginzburg-Landau free energy functional 
and defined at every lattice cell, 
whose coordinate we represent by $x$:
\begin{equation}
\label{eq:model}
   H[\{\phi_x\}] = -\frac{D}{2} \sum_{i,j}
\phi_x\Delta_{xy} \phi_y -\frac{r}{2}\sum_x^N  \phi_x^2
+\frac{g}{4}\sum_x^N \phi_x^4
\end{equation}
where $r>0$ and $g>0$ are the quadratic 
and quartic couplings of the GL 
theory while  the first term is proportional to the surface
energy.
We shall focus on the dynamical properties of the
GL model on the two dimensional deterministic Sierpinski Carpet (SC), 
of fractal dimension (Hausdorff dimension)
$d_f=\ln 8/\ln 3$. 

In order to construct the SC, one divides a square 
lattice of $L \times L$ cells,
with $L=3^n$, into $3 \times 3$ blocks of equal size and the 
cells contained
in the central block
are discarded. Dividing again each of the remaining blocks into 
$3 \times 3$ sub-blocks and discarding all the central elements
as many times as necessary to have 
the smallest sub-blocks constituted of a single cell,
one obtains a structure of $N$ cells, where $N = L^{d_f}$.

We make the assumption that
the evolution towards equilibrium of the order parameter, $ \phi_x$,
at the site $x$ is given by the Ginzburg-Landau equation:
\begin{equation}
\label{eq:Lang}
   \frac{\partial\phi_x(t)}{\partial t} = 
               - \Gamma \frac{H[\phi(t)]}{\delta\phi_x(t)}  
+ \eta_x(t)
\end{equation}

 Here $\eta_x(t)$ represents a Gaussian white noise 
with zero average and variance : 
$$
<\eta_x(t) \eta_y(t')>= 2 T_f \Gamma
\delta_{xy} \delta(t-t')
$$
where $T_f$ is the temperature of the final equilibrium
state, $\Gamma$ is a kinetic coefficient and $\delta_{xy}$ the
Kronecker symbol.

By substituting eq. (\ref{eq:model}) into (\ref{eq:Lang}) we find that
$\phi_x$, at any time
after the quench, changes according to the equation: 
\begin{equation}
\label{eq:Langevin}
\frac{\partial \phi_x(t)}{\partial t} =
\Gamma[ D \Delta_{xy}\phi_y(t) +r\phi_x(t)-
g \phi_x^3(t)]+ \eta_x(t)
\end{equation}

We shall write $\Delta_{xy}$
as  a difference operator in analogy with the discrete representation
of the Laplacian on Euclidean lattices. 
Periodic boundary condition are assumed, unless explicitly stated.
The operator $\bbox{\Delta}$ is defined as:
\begin{eqnarray*}
	\Delta_{xy}=  & 1    & \ \ {\rm if} \ \  x,y \ \  {\rm are}\ \ {\rm 
nearest} \ \ {\rm neighbor}\ \ {\rm cells} \\
	\Delta_{xx} = & -Z_x & \\ 
	\Delta_{xy} = & 0    &  \rm{otherwise}.
\end{eqnarray*} 

where $2\leq Z_x \leq 4$ 
counts the number of nearest neighbors of the site $x$.

At  temperature $T_f=0$, the
free energy has two two equivalent minima
$\phi=\pm \sqrt{r/g}$.
In the case of the spherical model we showed that the critical
temperature vanishes when the spectral dimension is less than $2$,
while in the scalar case the critical temperature of a Sierpinski Carpet
is finite \cite{Ruiz}. 
In the following, we shall concentrate on the dynamical properties
for deep quenches $T_f<<T_c$.

\section{auxiliary field approach}

In this section the so called 
auxiliary field method \cite{Bray},  
\cite{Desiena} and \cite{mazenko}, which has provided 
new insight into the ordering dynamics of 
translationally invariant systems, will be applied 
to the non conserved order parameter
dynamics on the Sierpinski Carpet.

In this approach one replaces the physical field $\phi_x$
by an auxiliary field, $m_x$, which varies in a smoother fashion
across the interfaces and renders approximations feasible.

One chooses a non linear transformation from $\phi_x(t)$ to a new 
field $m_x(t)$  in such a way that the latter obeys to
an equation simpler than the original one.
If such equation is linear,
the statistical properties of $m_x$ are equivalent to those of free
gaussian fields and analytical work can be performed.
Following the presentation of De Siena and Zannetti \cite{Desiena},
one way of determining the transformation
is to require that the auxiliary field linearizes 
the local part of the original equation of evolution (i.e. a zero
dimensional version of eq. (\ref{eq:Langevin}), obtained by 
setting $D=0$). For convenience,
unless stated explicitly we shall assume the stiffness constant $D$ 
and the kinetic coefficient $\Gamma$ to be 1.
The auxiliary field $m_x$ is introduced via the mapping: 
\begin{equation}
\label{eq:mazenko}
\phi_x=\phi(m_x)=\frac{m_x} {[1+(g/r) m_x^2]^{1/2}}
\end{equation}
In order to obtain the
equation of motion for the auxiliary field, $m_x$, 
we need to consider the non local term:
\begin{equation}
\label{eq:emme}
\sum_{y}\Delta_{xy} \phi(m_y)=\sum_{<y>} \phi(m_y)-Z_x\phi(m_x)
\end{equation}
where the sum $\sum_{<y>}$ is restricted 
to the $Z_x$ nearest neighbors of the site $x$.
Assuming that $\phi_x$ is a slowing varying function of $m_x$, one can expand
the field at a nearest neighbor site $y$ in a Taylor series:
\begin{equation}
\label{eq:nn}
\phi(m_y)=\phi(m_x)+\phi'(m_x)(m_y-m_x)+\frac{1}{2}\phi''(m_x)
(m_y-m_x)^2+ H.O.T.
\end{equation}
having indicated with primes the derivatives of $\phi_x$ with respect to
$m_x$ and
neglected higher order terms (H.O.T.) in the expansion.
Collecting together the terms from the $Z_x$ nearest neighbors one obtains:
\begin{equation}
\label{eq:lapla}
\sum_y \Delta_{xy} \phi_y \simeq \phi'(m_x)
\sum_{y}\Delta_{xy} m_y+\frac{1}{2}\phi''(m_x) \sum_{<j>}(m_x-m_y)^2
\end{equation}
The equation of motion for the auxiliary field $m_x$ reads: 
\begin{equation}
\label{eq:moto}
\frac{\partial m_x(t)}{\partial t} =\sum_{y}\Delta_{xy} m_y+r m_x
 -\frac{1}{2} Q(m_x)\sum_{y}\Delta_{xy}(m_x-m_y)^2
\end{equation}

To obtain eq. (\ref{eq:moto}) we have used the identities:
\begin{equation}
\label{eq:ide}
m_x=-\frac{- \phi(m_x)+g/r \phi(m_x)^3}{\phi'(m_x)}
\end{equation}
and 
\begin{equation}
\label{eq:ide2}
\frac{\phi''(m_x)}{\phi'(m_x)}=-\frac{3 (g/r) m_x }
{[1+(g/r)m_x^2]}=-Q(m_x)
\end{equation}
To proceed further, we consider a mean-field like approximation for the 
last term  in eq.(\ref{eq:moto}):
\begin{equation}
\label{eq:approx}
\sum_{y}\Delta_{xy}(m_x-m_y)^2 \simeq
\frac{1}{N} \sum_{xy}\Delta_{xy}(m_x-m_y)^2=-2
\frac{1}{N} \sum_{xy}\Delta_{xy}m_x m_y
\end{equation}

Introducing the abbreviation:
\begin{equation}
\label{eq:bis}
D_0(t)=-\frac{1}{N} \sum_{xy}\Delta_{xy}m_x m_y
\end{equation}
the equation of evolution for the auxiliary field $m_x(t)$
reads:
\begin{equation}
\label{eq:moto2}
\frac{\partial m_x(t)}{\partial t} =\sum_{y}\Delta_{xy} m_y+r m_x
-Q(m_x)D_0(t).
\end{equation}

Upon neglecting the last term in eq. (\ref{eq:moto2}),
i.e. setting $Q(m)=0$, we recover the
Kawasaki, Yalabik and Gunton (KYG) theory \cite{KYG}.
Alternatively, we expand  the function 
$Q(m_x)\simeq 3(g/r)m_x$ to first order in the coupling 
constant $g$  and write \cite{Desiena}: 

\begin{equation}
\label{eq:moto3}
\frac{\partial m_x(t)}{\partial t} =\sum_{y}\Delta_{xy} m_y(t)+[r -3
\frac{g}{r} D_0(t)]m_x(t).
\end{equation}
The field $m_x$ thus evolves according to an equation similar to the
one found in the spherical model and in the Hartree-like approximation,
the non linear term being treated self-consistently. 

In order to obtain the properties of the field $m_x$ we 
consider the eigenvalue problem:
\begin{equation}
\label{eq:auto1}
-\Delta_{xy}
v_y^{\alpha}=\epsilon_{\alpha} v_x^{\alpha}
\end{equation}
where  $v_x^{\alpha}$ is the $x$-th component of 
the eigenvector associated with 
the eigenvalue $\epsilon_{\alpha}$ of the operator 
$\bbox{\Delta}$.
There is an eigenvalue $\epsilon_0=0$ associated with the 
uniform mode whose eigenvector has all elements equal.
The asymptotic behavior of the solution of eq. (\ref{eq:moto3})
depends on the distribution of the smallest eigenvalues. 
After expanding  the field $m_x$
as a linear superposition of modes of amplitude $\tilde{m_{\alpha}}$: 

\begin{equation}
\label{eq:auto1b}
m_x(t)=\sum_{\alpha=0}^{N-1} \tilde{m}_{\alpha}(t)v_x^
{\alpha}.
\end{equation}
one finds that each component evolves independently as:

\begin{equation}
\label{eq:auto2}
\tilde{m_{\alpha}}(t)=\tilde{m_{\alpha}}(0)\exp{[-
\epsilon_{\alpha}t+B(t)]}
\end{equation}
The quantity $B(t)$ must be calculated self-consistently from the 
governing equation (\ref{eq:moto3}):
\begin{equation}
\label{eq:auto3}
\frac{\partial B(t)}{\partial t} = r-3 \frac{g}{r}<D_0(t)>_0
\end{equation}
where the average $<\cdot>_0$ is over the initial conditions of the
field $m_x$. 
Using eq. (\ref{eq:bis}) and the 
eigenfunction expansion of $m_x(t)$ we compute $<D_0(t)>_0$ as:

\begin{equation}
\label{eq:auto4}
<D_0(t)>_0=\frac{1}{N}
\sum_{\alpha} \epsilon_{\alpha}|\tilde{m}_{\alpha}(0)|^2
\exp{[-2\epsilon_{\alpha}t+2 B(t)]}.
\end{equation}
For $N \to \infty$ a continuum density of states approximation 
$\rho(\epsilon)\simeq \rho_0 \epsilon^{d_s/2-1}$
is appropriate and one can write: 

\begin{equation}
\label{eq:d0}
<D_0(t)>_0=
A \int d \epsilon \epsilon \rho(\epsilon) e^{-2 \epsilon t+2 B(t)}
\end{equation}
where $\sqrt{A}$ is proportional to 
the amplitude of the fluctuations of the field $m_x$
at the instant $t=0$.
One ends with a closed equation for $B(t)$:
\begin{equation}
\label{eq:closed}
\frac{\partial B(t)}{\partial t} = r-3 \frac{g}{r}e^{2 B(t)}
A \int d \epsilon \epsilon \rho(\epsilon) e^{-2 \epsilon t}
\end{equation}

From eq. (\ref{eq:closed}) one sees that
the
quantity $B(t)$ must behave as 
\begin{equation}
\label{eq:d1}
B(t) \sim \frac{d_s+2}{4}\ln t
\end{equation}

Thus asymptotically the quantity $<D_0(t)>$ goes to a constant value
$r^2/(3g)$, while
the equal-time correlation function diverges as:
\begin{equation} 
\label{eq:equaltime}
\frac{1}{N}\sum_x^N <m_x(t)m_x(t)>_0 \sim t .
\end{equation}

It is possible, now, to compute the correlation function:
\begin{equation}
\label{eq:d2}
<m_x(t)m_y(t)>_0=A \int d \epsilon \rho(\epsilon) e^{-2 \epsilon t+2 B(t)}
v_x(\epsilon)v^{*}_y(\epsilon)
\end{equation}
A scaling form for the above quantity is (see \cite{Guyer}):
\begin{equation}
\label{eq:d3}
<m_x(t)m_y(t)>_0 \sim t \exp\{-(R/t^{1/d_w})^{\frac{d_w}{d_w-1}}\}
\end{equation}
where $R=|\bbox{r_x}-\bbox{r_y}|$ is the distance between the sites $x$ 
and $y$. The above correlation function yields the following average
value for the droplet size, $L$, as a function of time:
\begin{equation}
\label{eq:d4}
<L^2(t)> \propto t^{\frac{2}{d_w}}
\end{equation}

Knowing the evolution of the field $m_x(t)$ it is possible to 
determine the properties of the original field $\phi_x(t)$.
Recalling that $m_x$ has a Gaussian distribution at all times,
one writes:
\begin{equation}
\label{eq:gauss2}
P_m(m_x,t;m_y,t)=\frac{1}{Z_m}
\exp{ \{-\frac{1}{1-\gamma^2}
[\frac{m_x^2}{S_0(x)}+\frac{m_y^2}{S_0(y)}-2\gamma \frac{m_x m_y}
{[S_0(x)S_0(y)]^{1/2}]}\}} 
\end{equation}
where $Z_m$ represents  a normalization factor:
\begin{equation}
\label{eq:zeta}
Z_m\equiv2 \pi \sqrt{S_0(x)S_0(y)(1-\gamma^2)}
\end{equation}
with
\begin{equation}
\label{eq:s0}
S_0(x)\equiv<m_x^2(t)>_0
\end{equation}
and
\begin{equation}
\label{eq:g0}
G_0(x,y;t)\equiv<m_x(t)m_y(t)>_0
\end{equation}

\begin{equation}
\label{eq:gamma}
\gamma(x,y)
\equiv\frac{G_0(x,y;t)}{\sqrt{S_0(x) S_0(y)}}
\end{equation}

The average value of the original
field $\phi_x(t)$ can be calculated from:

\begin{equation}
\label{eq:media}
<\phi_x(t)>=\int dm_x \int dm_y P_m(m_x,t;m_y,t)\phi(m_x)
\end{equation}

while its correlator
$G(x,y;t) \equiv <\phi_x(t)\phi_y(t)>$ is given by:
\begin{equation}
\label{eq:gauss}
G(x,y;t)=\int dm_x \int dm_y P_m(m_x,t;m_y,t)\phi(m_x)\phi(m_y)
\end{equation}

Using the form of the distribution function 
$P_m$ and eq.(\ref{eq:gauss}) one
finds for the correlation
function of the $\phi$ field the result (see \cite{mazenko}):
\begin{equation}
\label{eq:d5a}
<\phi_x(t)\phi_y(t)>=\int dm_x \int dm_y P_m(m_x,t;m_y,t) sign(m_x) sign(m_y)=
\frac{2}{\pi} \phi_{coex}^2 
\sin^{-1}[\gamma(x,y)]
\end{equation}
Substituting eqs.(\ref{eq:equaltime}),(\ref{eq:d3}) and
(\ref{eq:gamma}) into eq. (\ref{eq:d5a}) one obtains:
\begin{equation}
\label{eq:d5}
<\phi_x(t)\phi_y(t)>= \frac{2}{\pi} \phi_{coex}^2
\sin^{-1}[\exp{-(R/t^{1/d_w})^{\frac{d_w}{d_w-1}}}]
\end{equation}
which predicts a smooth decay of the correlation at long
distances and the existence of dynamical scaling behavior in the late stage 
growth.

To check the above results 
we have solved the equations of evolution (\ref{eq:moto3})
numerically on a Sierpinski Carpet of size $243 \times 243$
and monitored the growth by measuring the fluctuation of the 
homogeneous component of the order parameter:
$$
 C(0,t)=\frac{1}{N} (\sum_x^N\phi_x(t))^2
$$
which is  the zero component of the
structure factor and grows in time, as shown in
fig. (\ref{fig:fig1}),  with an apparent exponent
$\nu=0.93$ in agreement with the spherical model result \cite{andalo}
which gives $\nu=d_s/2$, with $d_s=1.86$.
The morphology of
the field $\phi_x$ obtained
inserting the solution $m_x(t)$ into the non linear mapping
eq. (\ref{eq:mazenko}) is shown in figs.(\ref{fig:fig2}-\ref{fig:fig4}) 
and reveals the existence of
large droplets growing in time in a fashion similar to that observed
on compact supports, as if the fractality 
affects only the mass-to-size ratio of the domains, but 
does not suppress the diffusive motion of the walls. 
 In this sense the auxiliary field method agrees quite well with the
behavior of the spherical model. 

In the next section 
we shall  compare these findings with a direct simulation of the
Langevin equation.

\section{Numerical results}

We have investigated 
numerically the non conserved dynamics on the Sierpinski 
Carpet starting from a disordered state, generated 
assigning to each cell 
a random number uniformly distributed in the interval $[-0.125,0.125]$
and assumed $r=g=1$.
In order to integrate numerically  eq. (\ref{eq:moto}) we adopted 
an Euler discretization scheme with time step $\Delta t=0.01$
and Sierpinski Carpets of different linear sizes 
 $L=27,81,243$ with periodic boundary conditions.
 We have checked that our results do not change appreciably
if we decrease further $\Delta t$.
We  also considered several temperatures quenches and runs up to
$t=1000$. The averages for the various quantities  presented below
refer to 50 independent random  initial configurations.
In order to measure the droplet size we have applied the Hoshen-Kopelman
algorithm \cite{Hoshen}
to label the individual clusters formed by nearest neighbor cells
characterized by the same sign of the order parameter. Quantitative 
measures of the droplet properties are their masses and the radii of gyration.
The mass, $M$,
is defined as the number of cell belonging to a droplet, while the 
radius of gyration is defined as:
$$
R^2=\frac{1}{M}
\sum_z^M[(x_z-x_{cm})^2+(y_z-y_{cm})^2]
$$ 
The sum is over the lattice sites belonging to a droplet, $(x_z,y_z)$ is
the position of the lattice sites $z$ and $(x_{cm},y_{cm})$
the position of its center of mass. To compute
the mean value of $R^2(t)$ as a function of time we
averaged over all droplets and over all initial configurations.

 The growth law for the average size of the
droplets is reported in fig.(\ref{fig:fig5})
and the data are compared with those referring to an
Euclidean square lattice of the same linear size. 
We observe that the growth is much slower in the first
system than in the second. In figs. ({\ref{fig:fig6}-{\ref{fig:fig8}) 
are shown  typical 
snapshots of the system:
the domain walls, separating opposite
phases, sit on locations where the surface energy cost is lower and
thermal fluctuations are needed to push the system out of these
minima.
Further evidence of the absence of power-law growth stems from the study of
$C(0,t)$, which is compared  in fig. (\ref{fig:fig9})
with the corresponding quantity in the Euclidean case,
known to grow as $t^{d/2}$. 

Since in the late stage
the characteristic width of the interfaces
does not change appreciably with time, 
another independent measure of the domain structure is given
by the ratio between the total number of sites 
and  the number of sites $N_p$ covered by interfaces,
the so called inverse perimeter density \cite{Rogers}:
$$
P(t)=N_{sites}/N_{p}
$$
By assigning a perimeter site every time the absolute value of $\phi$
was less than the value
$0.75$, one obtains an estimate of $P$ which is not too sensitive
to the above threshold. While $P(t)$ displays power law 
growth in translationally
invariant structures and grows proportionally to  $L(t)$,
in the fractal case saturates at a constant 
value for low temperature quenches as shown in fig.(\ref{fig:fig10}).

Another quantity with good self-averaging properties is
the surface energy density (fig.(\ref{fig:fig11})),
whose time dependence we studied and compared with the Euclidean 
case. For compact supports this quantity decays as  the inverse of
the domain size  $L^{-1}(t)$, 
while on the Carpet it displays slow relaxation and eventually freezes.

Based on these results
one is lead to the conclusion that, after the early regime,
the average domain size
and other indicators  do not grow with a power law behavior, 
as found in Euclidean lattices
and in the spherical model on the same fractal support. 
At zero temperature one finds a
breakdown of the self-similar dynamical scaling,
in contrast with the compact case.
Notice that the  crossover from a diffusive
and curvature driven 
dynamics to a thermally activated, Arrhenius-like,
dynamics is not disorder induced as in the case of
Ising diluted models  \cite{Huse}-\cite{puri}.
During the early stage the domains coarsen almost linearly with time 
because the curvature provides
the main driving mechanism, while at later times the growth
becomes much slower and eventually stops since the 
height of the barriers 
that the domain walls have to surmount 
increases with the domain size. The dynamics selects
configurations, in which  regions of opposite magnetization
are separated by boundaries
formed by a large  number of
voids. Since  these configurations are associated with low surface tension
the interfaces remain trapped.

 Interestingly, the crossover time from a power law
growth regime, during which the
trapping is not effective, to a frozen state is independent 
of the value of
the stiffness constant $D$. This is understood by recalling that
during the early stage the domain size grows diffusively as:
$$
R^2 \sim D t^{2/d_w}
$$
The pinning typically 
occurs when a moving interface encounters voids of size 
of the order of the interfacial width $w$, which scales as
$w \sim \sqrt{D}$. Recalling that the distance between two voids of
linear dimension $w$ is proportional to $w$ itself, (due to the 
deterministic nature of the fractal)
the typical separation of two domain walls is proportional to $w$.
The asymptotic radius of the droplets is:
$$
R_{\infty}^2 \sim w^2 \sim D
$$
which gives a crossover time $\tau$:
$$
D \tau^{2/d_w} \sim D
$$

Therefore the interfacial stiffness, $D$ determines 
the typical size of the droplets, but not the 
crossover time scale $\tau$ at which the freezing occurs.
 
Asymmetric, off-critical quenches lead to immediate equilibration
of one of the two phases, and fragmented patterns
are not observed. On the other hand,
when one considers the conserved order parameter dynamics on
percolation structures \cite{Castellano} even off-critical quenches 
fail to generate the ordered state during a cooling at reasonable rate
and lead to 
frozen dynamic behavior with interrupted growth at low temperatures.

{\bf Twisted boundary conditions.}

We finally looked at the effect  on the
phase ordering dynamics of a lattice with
twisted boundary conditions
along the x-direction of the lattice, maintaining 
the periodicity along the
y-direction . As 
$t \to \infty$ the
compact system orders and forms two large domains of
the opposite equilibrium phases, separated by  rather sharp interfaces.
 On the contrary, on a fractal
the system remains
frozen in a highly fragmented structure as shown in fig.(\ref{fig:fig12})
because of the presence
of voids, which reduce the role
of the curvature. To characterize the interfacial
roughness we consider the following correlator:
$$
G(x) \equiv \frac{1}{N_y}\sum_y \phi_(x,y)\phi(x,y)-(\frac{1}{N_y}
\sum_y \phi(x,y))^2
$$ 
where $N_y$ represents the width of the system and the two arguments denote
the longitudinal and transverse coordinate respectively.
In the case of the regular lattice, the peak of $G(x)$
narrows as the interface becomes straighter and sharper,
while in the case of the
fractal $G(x)$ remains multi-peaked, as the interface
pervades all the volume as illustrated in fig. (\ref{fig:fig13}).
 One can go further by defining a characteristic
width $W$ of the interface as:
$$
W^2 \equiv \frac{1}{N_x}\sum_x x^2 G(x) -(\frac{1}{N_x}\sum_x x G(x))^2
$$ 
As the time $t \to \infty$
$W(t)$ saturates to a much larger value than the
corresponding quantity in a Euclidean lattice with identical boundary 
conditions and random initial conditions.

\section{Discussion and conclusions}

On regular lattices, domain walls on sufficiently large
scales can be approximated in the continuum limit by smooth
curved interfaces, which either grow or shrink, whereas in the fractal
case the background lattice has a pronounced influence on the
phase separation.
 In a translationally invariant system the origin of the power law like
behavior
can be traced back to the interfacial motion: immediately after the 
temperature quench, the system relaxes rapidly along the directions with
the steepest gradients of the free energy 
and the order parameter sets
locally to one  its equilibrium values $\phi=\pm \sqrt{r/g}$, 
leaving the system 
disordered, due to the presence of multiple domains of opposite 
phases separated by thin walls.
 This early  regime is followed by a slower
dynamics during which the driving force stems
from  the movement of the domain walls in order to reduce the 
interfacial energy .
In other words, a curved portion of an interface will move
and its velocity will be proportional 
to its local curvature, while a 
planar portion  in the absence of thermal fluctuations 
comes at rest. The slow behavior of the late stage reflects 
the flatness of the energy landscape. 
 One can think of the representative point of the system 
as moving along a direction characterized by a very small
gradient of the free energy $H[\{\phi_x\}]$.

Let us consider the free energy Hessian :
$$
\frac{\delta^2 H}{\delta \phi_x \delta\phi_y}=
-\Delta_{xy} -[r  \phi_x-3 g \phi_x^2]\delta_{xy}
$$
its eigenvalue spectrum gives a measure of the stability of a 
selected configuration and reveals the origin of the slow directions 
of relaxation.
If at a given time the configuration is close to
one of the global minima of $H$, i.e.. those characterized
by uniform values of the field $\phi_x=\pm \sqrt{r/g}$, the spectrum
contains only positive eigenvalues and the 
resulting relaxation process 
is an exponential function of  time. On the other hand,
a  symmetric quench leads almost inevitably
to the formation of a domain structure,
which is associated to a saddle point and not 
to a minimum in phase space \cite{foot}.
Thus the Hessian calculated in correspondence of the droplet solution 
must have a negative eigenvalue which corresponds to the expanding
mode.
The approach to equilibrium consists of the passage from the unstable
droplet configuration to one of the absolute minima. Since the droplet 
size increases in time the free energy becomes flatter and flatter
along the unstable path and the instability is reduced.
Correspondingly the associated negative eigenvalue vanishes 
giving rise to a relaxation slower
than any exponential function \cite{foot2}.

After this premise, we can see how  the absence of translational invariance 
on a fractal support
modifies all the features of the spectrum of 
fluctuations described above.

\begin{itemize}
\item The bulk fluctuations are modified
and the system becomes more sensitive to thermal fluctuations
because the density of state exponent $d_s$
becomes smaller ( less than $d$, in fact).
However, these modes remain stable, since they correspond to positive
eigenvalues of the Hessian. 

\item 
The capillary wave spectrum, associated
with a kink solution, will also be modified 
and the zero mode Goldstone mode is suppressed by the following mechanism:
let us consider a rigid shift of a straight
interface parallel to one of the directions of higher symmetry of the
lattice. In fig. (\ref{fig:fig14}) we show the the multivalley
structure of the free
energy, obtained by changing the location of the wall.
A sequence of barriers hinders
the normal fluctuations of the wall which eventually 
serve to equilibrate the system in a standard lattice.
This rugged landscape will determine a slower growth of the domains
and  instead of a diffusive motion in a homogeneous space 
the interface  undergoes a diffusion in a
system of barriers of varying heights. 

\item
 The negative eigenvalue associated with the presence of
a curved interface remains negative only for droplets of small
size and becomes positive, when the curvature energy
becomes of the order of magnitude of the barriers.
\end{itemize}

Thus the role of the curvature energy, which provides the only 
driving force
in translationally invariant systems, is greatly reduced, 
because  the 
lacunarity tends to pin the interfaces
in configurations which locally  minimize the energy. The residual dynamics
can be understood in terms  of consecutive transitions among metastable
states, i.e. different basins of the free energy. Finally
at very low temperatures  
this mechanism eventually ceases and the dynamics becomes completely 
frozen.

Let us also remark that the multivalley structures is washed out
as the interfacial width exceeds the impurity size. 
Within the limit of infinite thickness
one should be able to recover the results of the spherical model. 
In this model, in fact, the width of the walls increases in time during the
ordering process; as a consequence one observes  scaling.

By contrast, in the scalar model
the interfacial width $W$ becomes asymptotically much shorter 
both than the domain size $L(t)$ and the
voids, so that the system fails to show power law growth.

The typical depth $B$ of the
local minima of the free energy can roughly be estimated to be proportional   
to the size of the domain $L$ to some positive power $\psi$.
The value $(d_f-1)$ is an upper estimate for the exponent $\psi$,
since the interfaces do not sit at random, but join
places where the surface cost decreases. 

Since the late stage growth proceeds only via thermally activated processes
one can roughly estimate the typical time $\tau$ for a droplet of size $L$
to overcome a barrier; this is given by the formula: 
$$
\tau=\exp(B/k_B T_f)\sim \exp[(L(t)^{\psi}/k_B T_f]
$$
Thus the domain size $L$ evolves logarithmically as
$$
L(t)\sim (k_B T_f \ln t)^{1/\psi}
$$
in agrrement with the droplet model \cite{Huse} and
stops completely at zero temperature because of the prefactor $T_f$.

To summarize,
we have studied a simple non random system and found that its late stage 
behavior is not characterized by power laws and is not
self similar , as in the case of 
of periodic substrates.  
We have also found that the treatment of the Ginzburg-Landau model
on fractal substrates
based on the auxiliary field method yields results which are consistent
with the solution of the spherical model on the same support,
but does not reproduce 
the phenomenon of pinning of domain walls which is observed numerically.
In fact, the spherical  model fails to reproduce 
a necessary feature  of phase separating systems,
i.e.  domain walls of finite thickness.
Moreover, contrary to the situation 
of the scalar model, within the spherical model
there is no activation energy 
associated 
with the creation of a kink 
since the energy gap between the ordered phase
and the instanton solutions, vanishes in the infinite volume limit.
This leads in the spherical model to a phase ordering
process which is diffusive and non-Arrhenius like. The fractality of the substrate
has the minor effect of rendering less effective the diffusive growth
(because $d_w>2$).
Finally the difference between the 
auxiliary field method and the numerical 
solution seems to be due to the
mean-field assumption eq. (\ref{eq:approx}), which 
underestimates the effects of the inhomogeneous lattice.

\section*{acknowledgements}
I am grateful to A. Petri with whom I have collaborated on related
problems, and to A. Maritan, C. Castellano, M. Zannetti 
and F. Sciortino for discussions and M. Natiello and A. Crisanti
for technical help.
I also thank A. Vulpiani for hospitality.  
Finally I wish to dedicate the
present work to the memory of Giovanni Paladin.

\begin{figure}[htpb]
\centerline{\psfig{figure=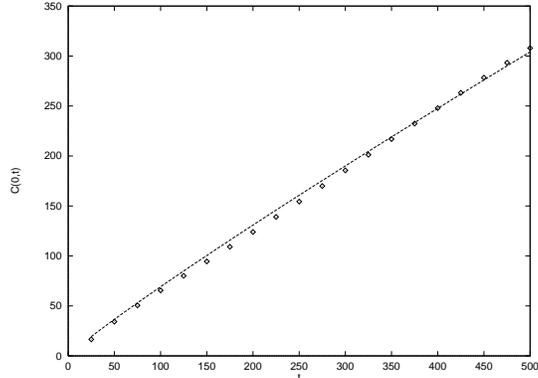,height=2in}}
\caption{Peak of $C(0,t)$ (measured in $r/g$ units), 
employing the
auxiliary field method, plotted versus time in arbitrary units (a.u.). 
The line represents 
the power law growth $t^{d_s/2}$, with $d_s=1.86$.} 
\label{fig:fig1}
\end{figure}

\begin{figure}[htbp]
\centerline{\psfig{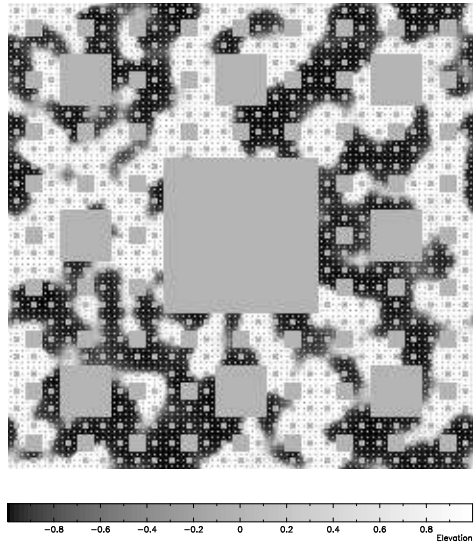}}
\caption{Morphology of the domains obtained by  solving numerically
the equations for the auxiliary field method on a lattice $243 \times 243$.
 Notice that the droplet structure
is not strongly affected by the presence of the holes, i.e. the
interfaces are not pinned.
The snapshot refers to times $t=25$ in the same arbitrary units
as fig.(1)}
\label{fig:fig2}
\end{figure}

\begin{figure}[htbp]
\centerline{\psfig{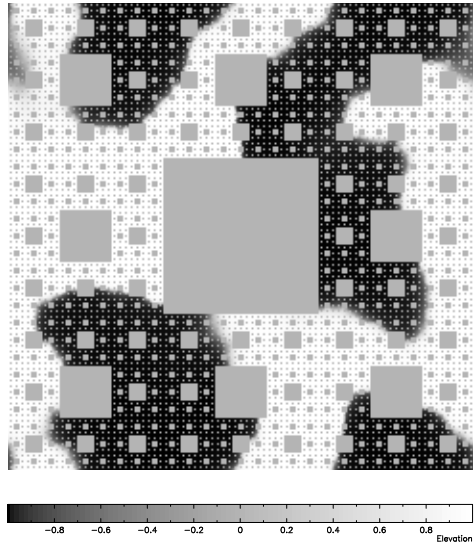}}
\label{fig:fig3}
\caption{Same as fig. (2), but at $t=250$ (a.u.)}
\end{figure}

\begin{figure}[htbp]
\centerline{\psfig{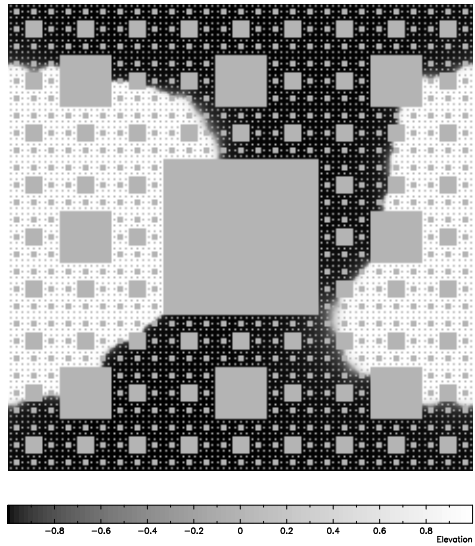}}
\caption{Configuration of the auxiliary field at  $t=2500$
(a.u.). }
\label{fig:fig4}
\end{figure}

\begin{figure}[htpb]
\centerline{\psfig{figure=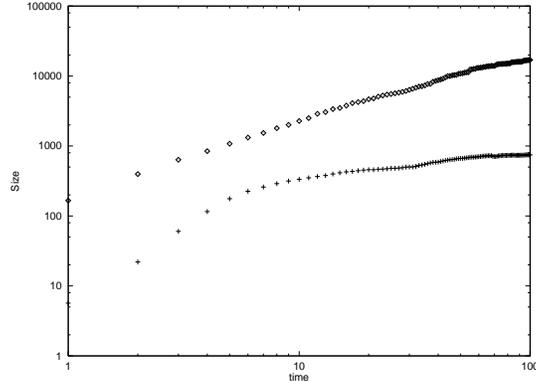,height=2in}}
\caption{Average size of the droplets (in lattice units)
versus time $t$ (a.u.)
calculated from the numerical solution of the Ginzburg-Landau equation.
The data represent  the averages over $50$  sets of random
initial conditions and noiseless dynamics $T=0$. The upper curve
refers to a periodic system $256 \times 256$, while the lower curve
to a Sierpinski Carpet of size $243$.  } 
\label{fig:fig5}
\end{figure}

\begin{figure}[htpb]
\centerline{\psfig{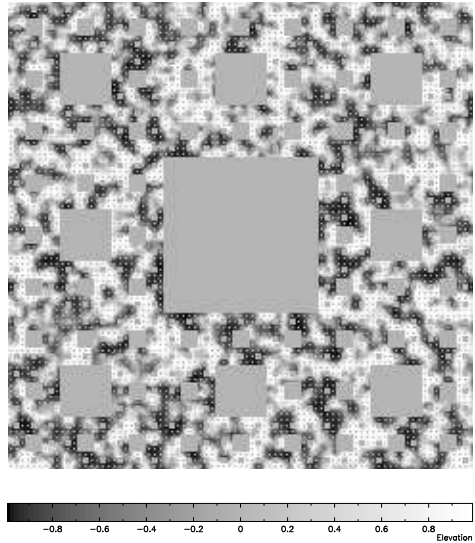}}

\caption{Instantaneous configurations at $t=10$ (a.u.) of the field
obtained solving numerically the Langevin equation 
on a Sierpinski Carpet $243 \times 243$.}
\label{fig:fig6}
\end{figure}

\begin{figure}[htpb]
\centerline{\psfig{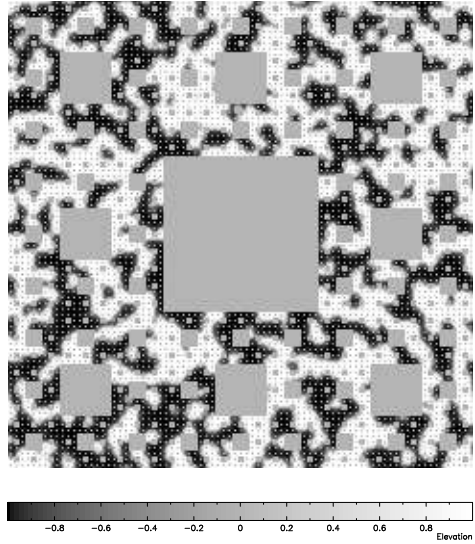}}
\caption{Same as fig. (6), but at $t=100$ (a.u.)}
\label{fig:fig7}
\end{figure}

\begin{figure}[htpb]
\centerline{\psfig{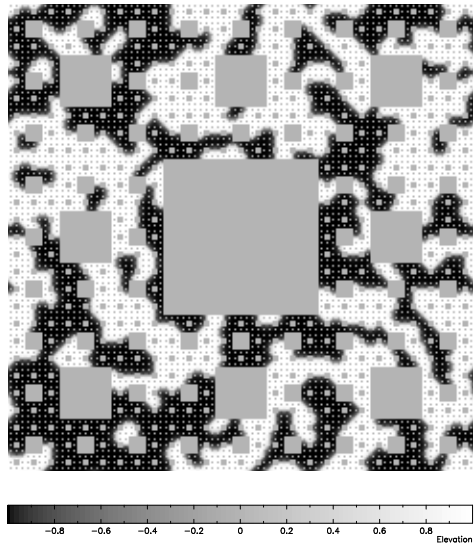}}
\label{fig:fig8}
\caption{Configuration of the scalar field at  $t=1000$
(a.u.). 
Notice that the walls remain frozen
at positions where the surface energy a local minimum.}
\end{figure}

\begin{figure}[htpb]
\centerline{\psfig{figure=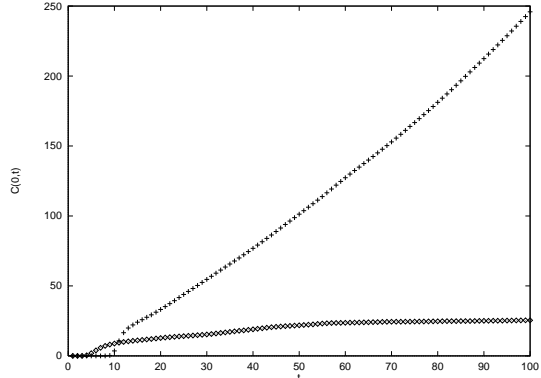,height=2in}}
\caption{Amplitude of the homogeneous component of the 
order parameter fluctuation versus time (a.u.) 
in the case of non conserved scalar order parameter dynamics on
a Sierpinski Carpet $243 \times 243$ (diamonds)
and  compact square 
lattice of the same size (crosses).} 
\label{fig:fig9}
\end{figure}

\begin{figure}[htpb]
\centerline{\psfig{figure=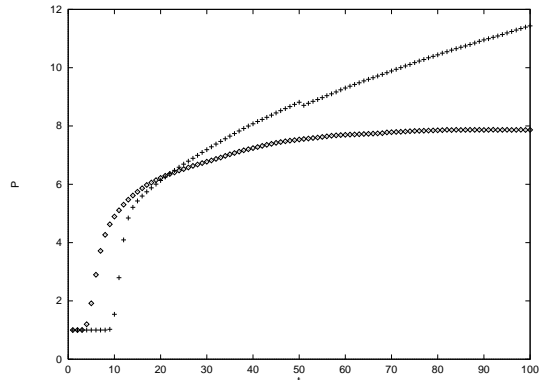,height=2in}}
\caption{Dimensionless inverse perimeter 
density versus time in arbitrary units for compact and fractal
supports. The symbols are the same as in fig. (9)}
\label{fig:fig10}
\end{figure}

\begin{figure}[htpb]
\centerline{\psfig{figure=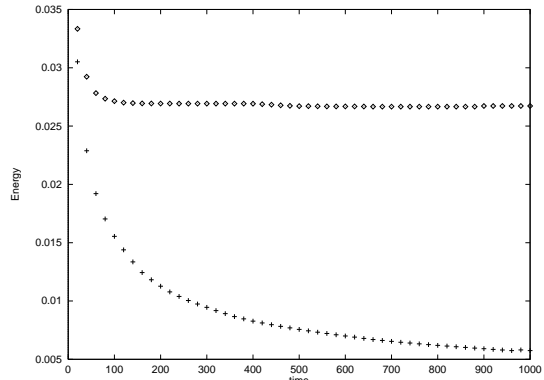,height=2in}}
\caption{Surface energy in arbitrary units
as a function of time for an Euclidean lattice
of size $256 \times 256$ and for a Sierpinski Carpet $243 \times 243$.
The symbols as in fig. (9)}
\label{fig:fig11}
\end{figure}

\begin{figure}[htpb]
\centerline{\psfig{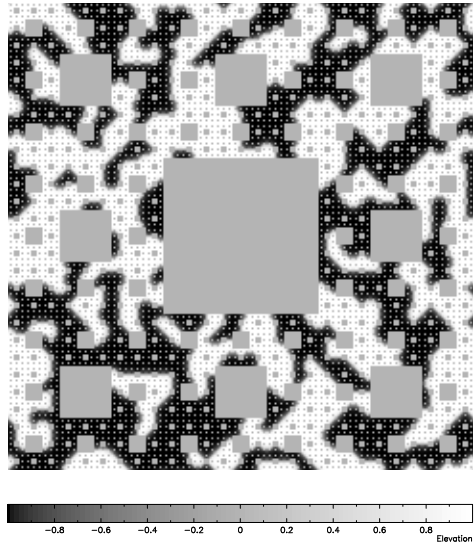}}
\caption{Typical configuration obtained in the presence of twisted boundary 
conditions along the vertical direction at $t=1000$ (a.u.)
for a Sierpinski Carpet $243 \times 243$.}
\label{fig:fig12}
\end{figure}

\begin{figure}[htpb]
\centerline{\psfig{figure=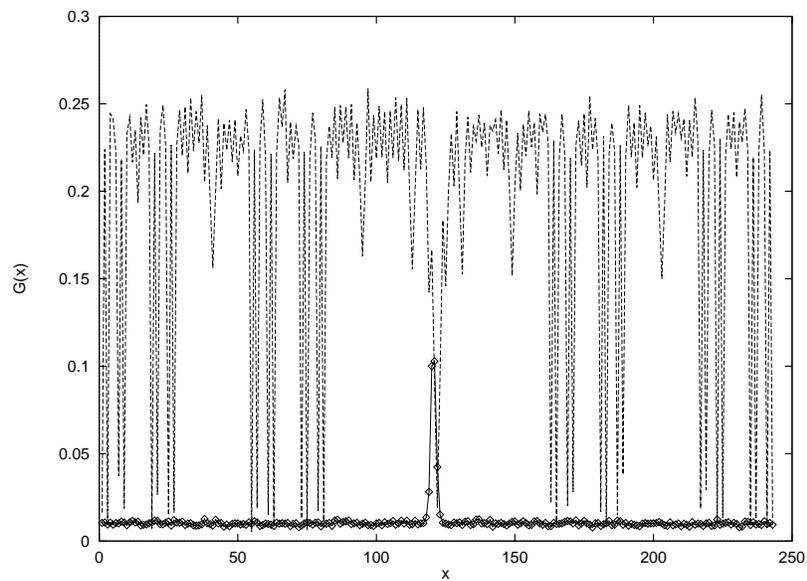,height=3in}}
\caption{Correlation $G(x)$ computed for 
for a standard lattice (lines and points) 
and for a Sierpinski Carpet (continuous line) with antiperiodic
boundary conditions along the x-direction at time $t=1000$ (a.u.). }
\label{fig:fig13}
\end{figure}

\begin{figure}[htpb]
\centerline{\psfig{figure=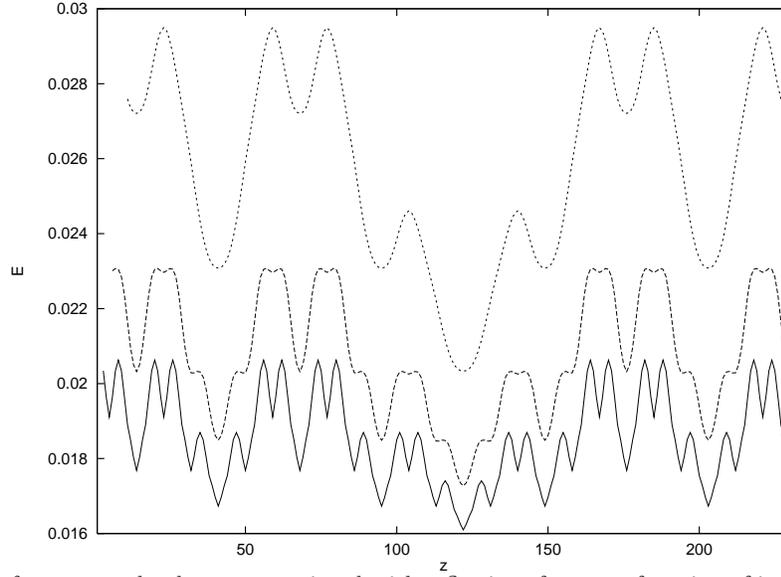,height=3in}}
\caption{
Surface energy landscape, 
associated with a flat interface as a function of its
location,
for three different choices of the width of the interface.
 The continuous curve refers to an interface
of width $w=4$ in lattice units, the intermediate 
curve refers to a curve of width
$w=8$, while the dotted  line refers to a diffuse 
($w=15$) interface. Notice the
disappearance of small valleys as the interfacial width increases.
The energy units are arbitrary and the x-axis is in lattice units. }
\label{fig:fig14}
\end{figure}


\begin{references}
\bibitem{rami}
A lattice is said to be finitely ramified if upon eliminating 
a finite number of lattice bonds one can isolate an arbitrary large 
subset of the infinite system.
\bibitem{foot}
To see that a droplet is a saddle point consider the fact that it
is stable with respect to bulk-like fluctuations, i.e. 
with respect to uniform changes of the field $\phi_x$,
but unstable with respect to curvature fluctuations.
\end{references}
\end{document}